# Symmetry-Driven Asynchronous Forwarding for Reliable Distributed Coordination in Toroidal Networks

Shenshen Luan, *Member, IEEE*, Yumo Tian, Xinyu Zhang, Qingwen Zhang, Tianheng Wang, Yan Yang, Shuguo Xie, *Member, IEEE*

*Abstract*—The proliferation of large-scale distributed systems — such as satellite constellations and high-performance computing clusters—demands robust communication primitives that maintain coordination under unreliable links. The torus topology, with its inherent rotational and reflection symmetries, is a prevalent architecture in these domains. However, conventional routing schemes suffer from substantial packet loss during control-plane synchronization after link failures. This paper introduces a symmetry-driven asynchronous forwarding mechanism that leverages the torus's geometric properties to achieve reliable packet delivery without control-plane coordination. We model packet flow using a topological potential gradient and demonstrate that symmetry-breaking failures naturally induce a reverse flow, which we harness for fault circumvention. We propose two local forwarding strategies — Reverse Flow with Counter-facing Priority (RF-CF) and Lateral-facing Priority (RF-LF) — that guarantee reachability to the destination via forward-flow phase transition points, without protocol modifications or additional in-packet overhead. Through percolation analysis and packet-level simulations on a 16×16 torus, we show that our mechanism reduces packet loss by up to 17.5% under a 1% link failure rate, with the RF-LF strategy contributing to 28% of successfully delivered packets. This work establishes a foundational link between topological symmetry and communication resilience, providing a lightweight, protocol-agnostic substrate for enhancing distributed systems.

## I. INTRODUCTION

THE next generation of intelligent systems is inherently distributed, relying on the cooperation of multiple agents to solve complex tasks. From high-performance computing of federated edge devices to coordinated control in quantum network and distributed sensing in satellite constellations, these systems require robust and efficient communication to achieve global objectives from local interactions and information exchange [1, 2]. The performance of core distributed algorithms — such as information broadcasting in satellite networks, entanglement distribution in quantum networks, and federated learning in high-performance computing environments—is critically dependent on the underlying communication network's ability to deliver messages reliably and with low latency [3, 4].

However, this foundation of cooperation is often undermined by real-world communication constraints. Limited bandwidth, quantization errors, dynamic topologies, and transient link failures are the norm rather than the exception in wireless settings [5]. These imperfections introduce packet loss and synchronization delays, which can severely degrade the convergence speed and accuracy of distributed algorithms [6]. A particularly pernicious challenge arises from the control-data plane synchronization in managed networks [7, 8]. Upon a link or node failure, the control plane (e.g., an SDN controller) must disseminate updated topology information, creating a synchronization interval during which the data plane's forwarding state is inconsistent. This desynchronization induces transient routing loops and blackholes, triggering substantial packet loss [9, 10]. Consequently, the network's ability to support fine-grained coordination is compromised, presenting a fundamental bottleneck for large-scale distributed intelligence.

The search for synchronization-free routing has spurred significant research. Convergence-free protocols like Failure-Carrying Packets (FCP) embed failure information in packet headers, enabling distributed path repair but incurring per-packet overhead [11]. Stateless data plane traversals (e.g., random walks) offer controller independence but often at the cost of significant path stretch and complex flow table management [12]. While these methods improve reachability, they often require specific protocol support or hardware, lacking the generality and lightweight properties desired for pervasive deployment in distributed systems [13, 14].Crucially, they fail to exploit the inherent geometric properties of structured network topologies to intrinsically mitigate the synchronization problem.

Manuscript received November 29, 2025; revised XX, XX, 2025; accepted XX, XX, 2026. This work was supported in part by National Natural Science Foundation of China under Grant 62401034. *(Corresponding author: Yan Yang and Shuguo Xie).* Shenshen Luan and Yumo Tian are co-first authors.

Shenshen Luan, Xinyu Zhang, Qingwen Zhang, Yan Yang, Shuguo Xie are with the School of Electronic and Information Engineering, Beihang University, Beijing 100083, China (e-mail: luanshenshen@buaa.edu.cn, by2302127@buaa.edu.cn, qingwenzhang@buaa.edu.cn, yanzi@buaa.edu.cn, xieshuguo@buaa.edu.cn). Shenshen Luan is also with National Excellent Engineer College of Beihang University, Beijing 100083, China.

Yumo Tian is with Hangzhou International Innovation Institute of Beihang University, Hangzhou 311100, China. (e-mail: tianyumo@buaa.eud.cn ).

Tianheng Wang is with Hefei Innovation Research Institute, Beihang University, Hefei 230012, China (e-mail: 12021041@buaa.edu.cn).

Code and data in this article are available online at https://github.com/FourierMax/Topology-Desynchronized-Forwarding

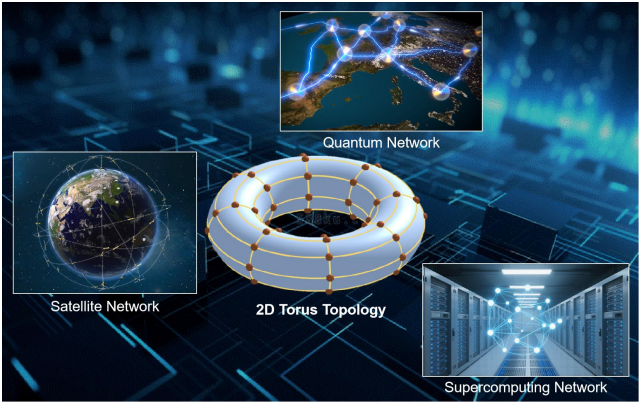

**Fig. 1.** Schematic of an 8×8 two-dimensional torus topology.

In parallel, structured network topologies, particularly the two-dimensional (2D) torus, have emerged as a foundational architecture for many modern distributed systems. Its high symmetry (translational, rotational, reflectional) and low diameter provide inherent robustness and efficient connectivity [15, 16]. This makes it the topology of choice for a diverse set of platforms: low-Earth-orbit (LEO) satellite constellations naturally form toroidal links via intra- and inter-plane connections [17, 18]; high-performance computing (HPC) clusters leverage it for low-latency communication [19, 20]; and quantum networks adopt it for scalable entanglement distribution [21, 22]. Despite its structural advantages, current routing schemes operating on toroidal networks have largely failed to exploit its geometric symmetries to intrinsically mitigate the synchronization problem.

*A. Contributions*

In this work, we bridge this gap by introducing a symmetry-driven, topology-desynchronized forwarding mechanism for 2D toroidal networks. We move beyond the conventional paradigm of strict control-data plane synchronization and reveal how the inherent rotational and reflection symmetries of the torus enable asynchronous, loss-tolerant packet forwarding. Our core contribution is a local, protocol-agnostic strategy that generates a reverse flow upon encountering a failure, leveraging the topology's symmetry to guide packets around faults without control-plane intervention or additional in-band signaling. The key innovations and contributions of this paper are summarized as follows:

- A Symmetry-based Analysis Framework for Torus Network: We model packet forwarding as a flow field defined by a topological potential gradient. Within this framework, we formally characterize how symmetry breaking due to failures naturally induces a reverse flow, and how the topology's rotational and reflection symmetries facilitate fault circumvention and flow stabilization.
- A Lightweight, Protocol-Agnostic Forwarding Mechanism: We design two practical, local forwarding strategies — Reverse Flow with Counter-facing Priority (RF-CF) and Lateral-facing Priority (RF-LF) — that operationalize this theory. They require no modification to packet headers, routing protocols, or control-plane interfaces, achieving reliability with zero control overhead.
- Theoretical Guarantees and Interdisciplinary Bridging: We establish the reachability of our mechanism, proving that packets are guaranteed to converge to a forward-flow phase transition point. Furthermore, we uncover a profound connection between the dynamic phase transition of our reverse-flow mechanism and the static structural phase transition of percolation theory, linking network dynamics to complex network physics.
- Comprehensive Validation with Implications for Distributed Cooperation: Through extensive packet-level simulations, we demonstrate a packet loss reduction of up to 17.5% in a 16×16 torus, with the RF-LF strategy contributing to 28% of successful deliveries. We contextualize these results, arguing that our mechanism provides a foundational communication primitive to enhance the resilience and efficiency of distributed learning, optimization, and inference tasks in satellite, HPC, and quantum networks.

The remainder of this paper is organized as follows: Section I reviews related works on fault-tolerant routing and highlights the gap in leveraging topological symmetry. Section II introduces our system model, defining the topological potential gradient and establishing the reachability guarantees under symmetry. Section III details the proposed reverse-flow forwarding strategies, RF-CF and RF-LF, along with the oscillation suppression mechanism. Section IV presents simulation results and performance analysis under bond and site percolation. Section V discusses the broader implications of our work, and Section VI concludes the paper. Through this structure, we aim to establish a foundational link between topological symmetry and communication resilience, offering a lightweight, protocol-agnostic solution that enhances the robustness of distributed optimization, learning, and inference tasks in bandwidth-limited and failure-prone environments.

*B. Related Works*

The efficacy of distributed optimization and learning algorithms hinges on timely state synchronization among participating nodes, which is critically vulnerable to link dynamics that prolong network convergence and degrade systemic performance [23, 24]. This synchronization interval induces transient routing inconsistencies—manifesting as transient loops and outage—that trigger substantial packet loss [9, 24]. Thus, control-data plane topology desynchronization emerges as the dominant contributor to packet loss [11, 13].

The search for synchronization-free routing has spurred significant research across multiple domains. Convergence-free routing protocols represent one major approach to addressing this challenge. Methods such as Failure-Carrying Packets (FCP) embed failure information in packet headers to enable distributed path recomputation, eliminating convergence delays but incurring per-packet overhead [11, 14]. Alternative approaches utilizing precomputed rules enable local multi-failure bypasses with minimal path stretch, yet they face fundamental constraints imposed by topological loop formation [25]. Various other data-plane mechanisms have been proposed to mitigate convergence delays, including randomized traversals [26] and metrics-based failover techniques [12]. However, these approaches often sacrifice path optimality or require specific hardware support, limiting their general applicability.

In the domain of deterministic routing theory, randomized routing guarantees k-1 failure delivery in k-connected graphs, though with potential path stretch, while deterministic arc-disjoint trees require strict k-edge-connectivity [23]. The Tunneling on Demand (TOD) method achieves near-complete protection through minimal labeling in arbitrary topologies [27], revealing inherent resilience-overhead trade-offs that characterize many fault-tolerance schemes.

For lattice/2D torus topologies, the relationship between characteristics such as routing paths, network capacity, and network scale has been theoretically investigated [28]. Fundamental advances in 2D torus fault-tolerant routing theory address critical graph-theoretic constraints. Rabin diameter equals $d + 1$, providing optimal-length node-to-node paths under $2r - 1$ node faults via $O(r^2)$ algorithms [29]. Fault model to clusters (diameter $\leq 1$) is proved with tolerance to $2n - 1$ clusters while maintaining paths of length $O(nk)$ [30]. Deadlock constraints on 2D torus can be resolved through Bubble flow control, enabling fully adaptive minimal routing without virtual channels [31]. Collectively, these works relax Menger's theorem [32] limitations for toroidal graphs while preserving path optimality and polynomial complexity. Crucially, packet loss remains an inevitable byproduct of topology synchronization itself.

Despite these substantial advances, a critical limitation persists across existing solutions: they largely fail to exploit the inherent geometric properties of network topologies. The fundamental challenge of packet loss induced by topology synchronization itself remains inadequately addressed. Current approaches either depend on additional protocol support, introduce substantial overhead, or require maintaining specific network states. There exists a notable absence of lightweight, fully distributed fault-tolerance mechanisms that leverage the intrinsic symmetric properties of topologies like the torus to achieve zero-control-overhead operation. Our work directly addresses this research gap by exploring how the rotational and reflection symmetries of toroidal networks can be harnessed to intrinsically facilitate fault circumvention, without requiring packet modifications, protocol changes, or reliance on control-plane intervention.

## II. SYSTEM MODEL

### A. Flow Field Defined by Topological Potential

To harness the symmetry of the torus, we model packet forwarding as a flow within a scalar potential field. Define the topological potential $\Phi_d(v)$ for a node $v$ with respect to a destination node $d$, as the length (hop count) of the shortest path from $v$ to $d$ in the original, fault-free torus network $G_0$, forming a gradient field that guides packet forwarding — a concept with roots in potential-based routing schemes [33] but here specialized to exploit the symmetries of the torus.

For any destination node, this potential resides at the node itself, with equal-potential lines forming concentric rings. Flows adhering strictly to the shortest-path criterion exhibit monotonically decreasing potential. A forward flow is characterized by decreasing potential with increasing hop count. Conversely, a reverse flow exhibits non-decreasing potential. Nodes connecting reverse to forward flows are termed flow phase transition points. Due to translational and reflection symmetries of the torus topology, four distinct forward flow fields exist for any destination node, geometrically centered on it as shown in Fig. 2. Boundary nodes of these fields constitute flow phase transition points.

Under fault-free conditions, flows complete shortest-path forwarding within their respective forward flow fields. During faults, reverse flows, as per the proposed method (Method section), alter packet topological potential to adjust forward flow paths. The Lateral Forwarding Priority strategy uses lateral reverse flows to switch forward flows within the packet's originating flow field. The Opposite Forwarding Priority strategy employs opposite reverse flows to direct packets to other forward flow fields. The Priority Strategy Reversal mechanism prevents packet oscillation within a single flow field, ensuring convergence to a forward flow phase transition point.

### B. Forward-Reverse Flow Phase Transition

Based on the torus topology's symmetry, the reverse flow forwarding mechanism guarantees packet convergence to a forward flow phase transition point. Moreover, at least one forward flow path connects to any destination node. Let $G = (V, E)$ be a connected subgraph of a toroidal network after arbitrary link/node failures, and let $d \in V$ be any destination node. Suppose routing tables are precomputed on the original fault-free topology using shortest-path algorithms. Then there exists a finite path $P = (v_k \to v_{k-1} \to \cdots v_0)$ with $v_0 = d$ and $k > 1$, such that every forwarding step $v_i \to v_{i-1}$ along P is a forward flow. Since G is connected and $|V| \geq 2$, the destination d has at least one neighbor u such that the edge $(u, d) \in E$ is operational. By construction of the routing table on the fault-free topology, the next-hop from u to d is directly through the interface connecting to d. Given edge $(u, d)$ is operational, u forwards packets to d via this interface without detecting a failure, satisfying the forward flow condition. Thus, the path $P_1 = (u \to d)$ exists. Iteratively, for each subsequent node $v_i$ along the route to d, if all edge $(v_i \to v_{i-1})$ are operational, the entire path $P = (v_k \to v_{k-1} \to \cdots v_0)$ consists of forward flow steps. Repeating the argument on this subpath constructs a valid forward flow path to d. Minimal path length is 1 (e.g., $u \to d$), and maximal length is bounded by $\text{diam}(G)$.

> REPLACE THIS LINE WITH YOUR MANUSCRIPT ID NUMBER (DOUBLE-CLICK HERE TO EDIT) <     4</->

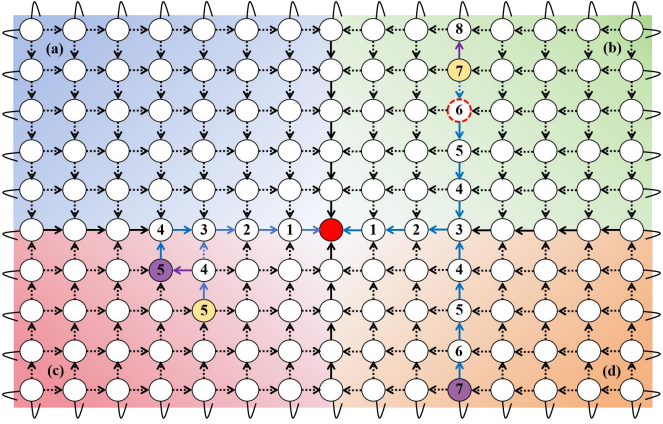

**Fig. 2.** Topological potential and flow fields. The torus topology is unfolded into a plane centered on the destination node and divided into four flow fields (a)–(d) according to topological potential. The numbers on the nodes indicate each node's topological potential. Darker colors denote larger potential. There are two data flows forwarded with the assistance of reverse flow, in which the source node is painted yellow respectfully and the destination node is red. The forward-flow phase-transition point is drawn purple, forward-flow paths are blue, and reverse-flow paths are purple. Flow field (c) shows a reverse flow induced by a link failure (blue dashed arrow). Flow field (b) shows a reverse flow caused by a node failure (red dashed node) that crosses into flow field (d) for forwarding.

*C. Rigorous Proof of Forward Flow Reachability*

Let the original failure-free two-dimensional toroidal network be a finite undirected graph $G_0=(V, E)$, where V is the set of nodes and E is the set of edges.

Let $F \subseteq E$ be a set of failed edges (node failures can be modeled as the failure of all their incident edges). The post-failure network is the subgraph $G_1=(V_1, E_1)$, where $E_1 = E \setminus F$, and $V_1$ is the set of nodes incident to edges in $E_1$. We assume $G_1$ is connected.

Fix a destination node $d \in V_1$. For each node $v \in V$, its forwarding table is generated based on $G_0$ using a shortest path algorithm. Specifically, $\forall v \in V$, define the function $next_d(v)$ as the unique next-hop neighbor from $v$ to $d$ in $G_0$ (if the shortest path is unique), or one selected according to some deterministic rule (e.g., smallest port ID) from multiple equal-cost next-hops. This function remains unchanged before and after failures.

Define the topological potential function $\Phi_d: V \to N_0$, where $\Phi_d(v)$ is the length (hop count) of the shortest path from v to d in $G_0$. Clearly, $\Phi_d(d) = 0$. An edge $(u, w) \in E_1$ is called a forward edge from u if and only if $\Phi_d(w) < \Phi_d(u)$. A path $P = (v_0, v_1, ..., v_k)$ is called a forward path if and only if $\forall i \in [0, k-1]$, $(v_i, v_{i+1})$ is a forward edge and $\Phi_d(v_{i+1}) < \Phi_d(v_i)$.

**[Theorem of Forward Flow Reachability]** Assume the post-failure subgraph $G_1$ is connected and the destination node $d \in V_1$. Then for any source node $s \in V_1$, there exists a forward path from s to d in $G_1$.

**Proof:**

Let $\Phi_d(s) = k$. If $k=0$, then $s=d$. The path $P=(d)$ is a trivial (length 0) forward path. If k=1, then s is a direct neighbor of d in $G_0$. By the definition of the forwarding table, $next_d(s)=d$. If the edge $(s, d) \in E_1$ (i.e., the edge is operational), then the path $P=(s, d)$ satisfies $\Phi_d(d)=0<1=\Phi_d(s)$ and is a forward path. If $(s, d) \notin E_1$, we require the following lemma to guarantee the existence of an alternative forward path.

**Lemma 1.** Let $G_1$ be connected, $d \in V_1$, and $s \in V_1$ such that $\Phi_d(s)=k>0$. Then there exists a node $w \in V_1$ such that $(s, w) \in E_1$ and $\Phi_d(w)<k$.

**Proof of Lemma 1:** Assume, for the sake of contradiction, that all alive neighbors $w$ of $s$ satisfy $\Phi_d(w) \geq \Phi_d(s)=k$. Consider a shortest path $P_{s\text{-}d}$ from $s$ to $d$ in $G_0$. The first edge $(s, next_d(s))$ of this path must satisfy $\Phi_d(next_d(s)) = k-1$. By our assumption, this edge has failed, i.e., $(s, next_d(s)) \notin E_1$.

Now consider any path P' from $s$ to $d$ in $G_1$ (such a path exists because $G_1$ is connected). Path P' must start from s and go through one of its alive neighbors $w_1$. By our contradictory assumption, $\Phi_d(w_1) \geq k$. Define the function $L: V_1 \to N_0$ as the length of the shortest path from node $v$ to $d$ in $G_1$. Clearly, $L(d) = 0$ and $L(v) \geq \Phi_d(v)$ for all $v \in V_1$, since $G_1$ is a subgraph of $G_0$.

Let $w_1$ be the next node after s on an actual shortest path from $s$ to $d$ in $G_1$. It follows that:

$$L(w_1) = L(s) - 1 \geq \Phi_d(s) - 1 = k - 1 \quad (1)$$

Furthermore, since $w_1$ is a neighbor of s and $(s, w_1) \in E_1 \subset E$, the definition of $\Phi_d$ in $G_0$ implies:

$$\Phi_d(w_1) \geq \Phi_d(s) - 1 = k-1 \quad (2)$$

Combining (1) and (2), and since $L(w_1) \geq \Phi_d(w_1)$, we conclude that $L(w_1)=\Phi_d(w_1)=k-1$. This means $\Phi_d(w_1) < \Phi_d(s)=k$, and the edge $(s, w_1)$ is alive, which contradicts the initial assumption. Therefore, Lemma 1 holds.

Assume that for all nodes $u \in V_1$ satisfying $\Phi_d(u) = m$ (where $0 \leq m \leq n$), there exists a forward path from u to d in $G_1$.

Now consider any node $s \in V_1$ such that $\Phi_d(s) = n+1$.

By Lemma 1, there exists a node $w \in V_1$ such that $(s, w) \in E_1$ and $\Phi_d(w) < n+1$, i.e., $\Phi_d(w) \leq n$. By the induction hypothesis (IH), for this w (since $\Phi_d(w) \leq n$), there exists a forward path $P_w$ from w to d. Therefore, the path $P=(s) \oplus (w) \oplus P_w$ (where $\oplus$ denotes path concatenation) is a path from s to d. Its first hop $(s, w)$ satisfies $\Phi_d(w) < \Phi_d(s)$, thus it is a forward edge; the remainder $P_w$ is a forward path by the IH. Hence, P is a forward path. By the principle of mathematical induction, the theorem is proven.

This proof, starting from a formal model and utilizing the relationship between the actual path length function L(v) in $G_1$ and the topological potential function $\Phi_d(v)$ defined on $G_0$, rigorously ensures the feasibility of the inductive step through the introduction of Lemma 1. It demonstrates that even if the primary next-hop link fails, network connectivity guarantees the existence of another alive link along which the topological potential necessarily decreases. This provides a solid

> REPLACE THIS LINE WITH YOUR MANUSCRIPT ID NUMBER (DOUBLE-CLICK HERE TO EDIT) <

mathematical guarantee for the core concept that the reverse flow mechanism is merely a temporary deviation, which will eventually return to a forward flow and reach the destination. It fundamentally establishes the correctness and reliability of the proposed method.

III. PROPOSED METHOD

*A. Reverse Flow Generation and Transmission*

During packet flow forwarding in a toroidal topology network, a forwarding node retrieves the destination address and corresponding egress interface from its routing table. Prior to forwarding, the node verifies the operational status of this egress interface, ascertained via hardware drivers or link detection protocols. If operational, the flow transmits normally via the egress buffer. If non-operational, the node selects an alternative egress interface, generating reverse flow. This generation operates independently of the control plane or routing policies, relying solely on real-time forwarding-plane link status, constituting the core mechanism for topology-desynchronized forwarding. Based on node degree, reverse flow employs two forwarding strategies introducing stochasticity for varied connectivity: Counter-facing Priority Strategy and Lateral-facing Priority Strategy (Pseudocode in Appendix A).

Counter-facing Priority Strategy prioritizes the diametrically opposite (counter-facing) interface for transmission. For degree-2 nodes, this is the sole alternative; successful transmission generates counter-reverse flow, failure causes discard. For degree-4 nodes, counter-facing interface priority applies; success generates counter-reverse flow. Failure triggers sequential attempts at clockwise then counter-clockwise lateral interfaces; lateral success generates lateral-reverse flow, dual lateral failure causes discard.

Lateral-facing Priority Strategy is inapplicable for degree-2 nodes. For degree-4 nodes, clockwise or counter-clockwise lateral interface priority applies; success generates lateral-reverse flow. Dual lateral failure triggers counter-facing attempt; success generates counter-reverse flow, failure causes discard. Reverse flow generation involves no flow marking or modification.

*B. Reverse Flow Forwarding and Annihilation*

Supposing node A forwards generated reverse flow using its originating strategy (Counter-facing or Lateral-facing Priority). Upon reaching node B via the new egress interface, node B identifies reverse flow status: it records the ingress interface, determines the standard egress interface via routing table lookup (destination address), and compares these interfaces. Identical interfaces classify the flow as reverse flow, triggering specialized forwarding; differing interfaces enable standard forwarding.

Specialized reverse flow forwarding at node B: For degree-2 nodes, forward via the counter-facing interface relative to ingress; failure triggers loopback via ingress. For degree-4 nodes executing Counter-facing Priority strategy, prioritize counter-facing forwarding; failure continues Counter-facing Priority logic. For degree-4 nodes executing Lateral-facing Priority strategy, prioritize lateral-facing forwarding; failure continues Lateral-facing Priority logic.

Due to 2D torus translational symmetry, reverse flow inevitably reaches a node N where the standard egress interface (routing table derived) differs from the ingress interface. Here, the flow reverts to forward flow under standard routing logic, achieving reverse flow annihilation as shown in Fig. 3 (Pseudocode in Appendix A).

*C. Oscillation Suppression Strategy*

Continuous application of a single priority strategy during reverse flow forwarding may induce localized oscillation. Mitigation requires strategy switching based on accumulated hop count. A Strategy Switch Threshold (SST) is defined as twice the geodesic node count. During forwarding: if hop count $\leq$ SST, apply the original priority strategy; if hop count > SST, apply the opposite strategy. In circular topologies (torus fundamental units), partition into disjoint arcs causes hop-count expiry discard; connected arcs enable oscillation escape via strategy switching. The circular topology is the fundamental unit of the torus topology; the above processes and conclusions naturally extend to the torus topology (Pseudocode in Appendix A).

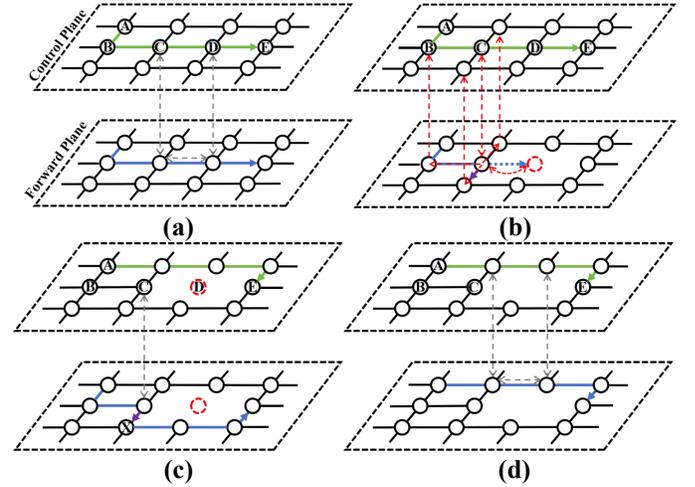

**Fig. 3.** Symmetry breaking between the control plane and the forwarding plane and asynchronous forwarding. (a) Under normal operation, the control plane and the forwarding plane achieve topological symmetry via link probing. Routing is in a converged state and forwarding proceeds normally. (b) When node D fails, due to the delay of the link-probing protocol, symmetry between the control-plane and forwarding-plane topologies is broken. The node-failure information is flooded between the control plane and the forwarding plane through multiple exchanges to synchronize the routing information databases. During route recomputation, packet loss occurs at node C. However, the reverse-flow mechanism forwards packets directly to adjacent nodes along the direction of the purple arrows without computing routes. (c) During routing reconvergence, the reverse-flow mechanism forms a temporary forwarding path, producing asynchronous

forwarding with respect to the control plane, as shown by the green and blue paths. The purple segment within the blue path is the reverse flow, and node X is the forward-flow phase-transition point. (d) After routing reconvergence completes, the forwarding-plane topology is again symmetric with the control-plane topology. The forwarding plane resumes shortest-path forwarding according to the control-plane routes, restoring synchronized routing and forwarding.

## IV. Performance Evolution

Method efficacy was validated via MATLAB discrete event-driven packet-level simulation. Metrics included packet loss rate, maximum path length (hops), and reverse flow utilization proportion under bond/site percolation.

Compared methods: Normal Flow (NF), Loop-Free Alternate (LFA), Proposed (Reverse Flow with Counter-facing /RF-CF, Reverse Flow with Lateral-facing/RF-LF). Topology: 16×16 torus.

Failures followed an exponential distribution across three regimes: Low probability (0.0001–0.01), Medium probability (0.001–0.1), High probability (0.01–1).

Traffic simulation: A random source node generated and sent one packet per second to a random destination via shortest-path routing. 1000 independent simulation replicates per parameter configuration ensured the minimum error of the results.

### A. Forwarding Performance under Percolation

**Packet-loss reduction performance.** By using reverse flow to bypass faulty nodes and then returning to forward flow via topological potential, data flows can be forwarded under topological desynchronization, which theoretically reduces packet loss. To verify this conclusion, we examined the packet-loss performance of the reverse-flow forwarding mechanism under bond and site percolation [34] (Detailed results of different topology are shown in Appendix Fig. B-1 and Fig. B-2). A random pair of nodes in a 16 × 16 torus topology was selected as the source and destination for packet transmission. By inducing link failures with varying probabilities, we recorded packet-loss rates for different methods via Monte Carlo trials. Fig. 4 presents the raw and fitted packet-loss data for the different forwarding methods. The red curve denotes the packet-loss rate of the normal flow (no-failure routing strategy), serving as the benchmark for comparison. Under bond percolation, the LFA, RF-CF (counter-flow priority) and RF-LF (lateral-flow priority) reverse-flow forwarding mechanisms all effectively reduce packet loss. The lateral-flow-priority reverse-flow mechanism achieves the best packet-loss suppression, reaching up to 17.5% improvement over NF under bond percolation and up to 11.2% under site percolation.

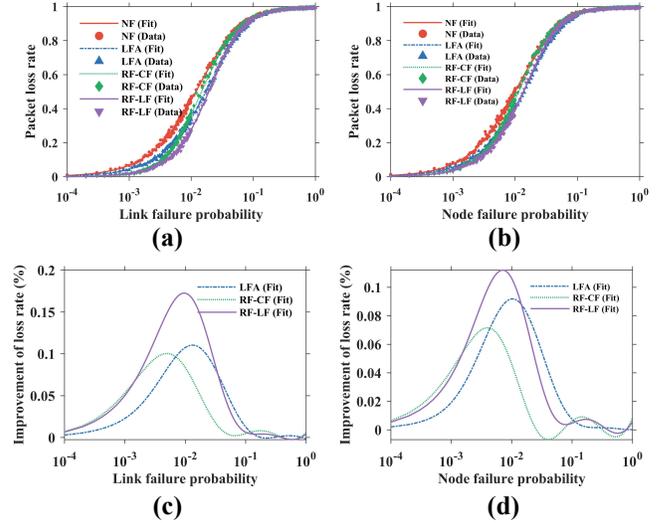

**Fig. 4.** Packet-loss rate improvement comparison. Packet-loss performance of three methods are shown under random link failures (bond percolation) and random node failures (site percolation). (a) and (b) show absolute packet-loss rates under bond and site percolation; panels (c) and (d) show the improvement of the LFA, RF-CF, and RF-LF methods relative to the no-failure routing strategy.

**Trade-off of maximum hop count.** The reverse-flow mechanism forwards data by traversing against the topological potential, which increases packet hop count. Fig. 5 shows how the maximum hop count of packets successfully delivered to the destination varies with the failure rate. Compared to the NF method, LFA strategies slowed down the rate of decline of the maximum hop count. Under reverse-flow strategies, when the failure probability is small ($P<P_c$), reverse flow can effectively bypass failure locations and achieve successful forwarding, causing the maximum hop count to increase with failure rate and peak at $P_c$ (bond percolation: RF-CF 0.48%, RF-LF 0.57%; site percolation: RF-CF 0.34%, RF-LF 0.40%). As the failure rate further increases, network fragmentation raises the probability of isolated islands. Reverse flow can no longer deliver packets to the destination. The maximum hop count then declines with increasing failure probability until it converges with the NF and LFA strategies. These results indicate that the proposed reverse-flow mechanism can achieve over 10% packet-loss suppression with a maximum hop-count increase of no more than 3 hops under bond percolation and no more than 2 hops under site percolation.

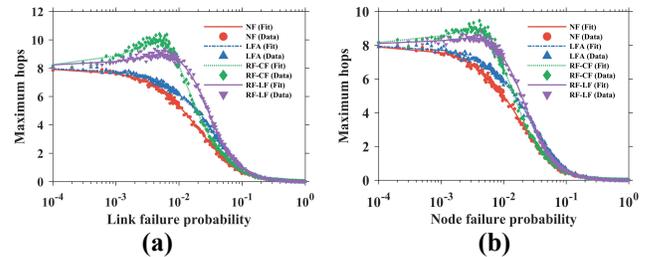

**Fig. 5.** Maximum hop-count variation. The RF mechanism detour packets, causing the network hop count increasing. a Under bond percolation, the hop count increases from 8 hops to a maximum of approximately 10.6 hops. b Under site

percolation, the hop count increases from 8 hops to a maximum of approximately 9.5 hops.

*B. Contribution of Reverse Flow Mechanism*

We define the proportion of packets that were successfully forwarded to the destination via the reverse-flow mechanism among all packets that arrived successfully as the RF packet ratio, which quantifies the contribution of reverse flow to reliable delivery. Fig. 6 shows that this mechanism effectively improves delivery success under both bond and site percolation. In bond percolation, packets successfully forwarded via the RF-LF mechanism account for up to approximately 28% of all successfully delivered packets, while packets forwarded via the RF-CR mechanism account for up to approximately 14% of all successfully delivered packets. In site percolation, the corresponding values are approximately 18% and 11%, respectively. Similarly, we compute the total number of hops of all successfully delivered packets and calculate the proportion of those hops that were taken via the reverse-flow mechanism, defined as the RF hops ratio. In contrast to the RF packet ratio, RF-CF exhibits a higher RF hops ratio than RF-LF. This indicates that RF-LF primarily acts as a connecting flow between forward flows during packet forwarding, whereas RF-CF primarily serves to deliver packets to their destinations.

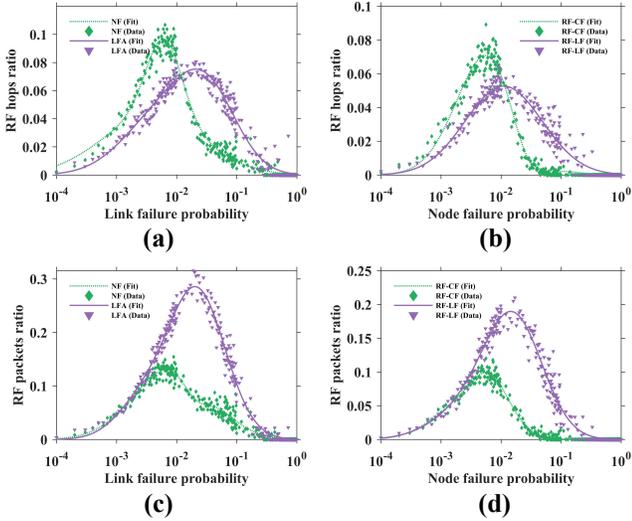

**Fig. 6.** Contribution of reverse flow to reliable forwarding. The proportion of hops and the proportion of packets attributable to the reverse-flow mechanism among packets successfully delivered to the destination are measured. a and b show the contribution of reverse flow to hop counts in successfully delivered packets under bond and site percolation, respectively. c and d show, under bond percolation and site percolation respectively, the proportion of packets successfully delivered to the destination via the reverse-flow mechanism among all packets that were successfully delivered to the destination.

*C. Scalability Analysis*

To evaluate the scalability of our proposed mechanism, we extended our simulations to torus networks of varying sizes, from 8×8 to 16×16. The complete results under both bond and site percolation are detailed in Appendix B (Figs. B-1 and B-2). The key finding is that the performance improvements observed in the 16×16 torus are consistent and generalizable. Specifically, the RF-LF strategy consistently achieves the highest packet loss reduction across all network scales. Furthermore, the contribution of the reverse-flow mechanism (quantified by the RF packet ratio) remains significant, typically between 15% and 30%, indicating its robust role in reliable delivery irrespective of network size. The performance gains are more pronounced in smaller topologies, which can be attributed to their higher sensitivity to individual link failures due to a smaller average path length. These results collectively confirm that our symmetry-driven forwarding mechanism is not an artifact of a specific network scale but provides a scalable, lightweight solution for toroidal networks.

## V. DISCUSSION

While classical percolation theory primarily characterizes the connectivity phase transition of static networks using the size of the largest connected component (P∞) as the order parameter, our work reveals a profound and previously unexplored connection between this structural transition and the dynamic behavioral transition observed in packet forwarding within a toroidal topology. The max hops undergoes a continuous transition near the critical failure rate $P_c$, signifying the work of efficient, symmetry-enabled fault tolerance. This suggests that the reverse-flow mechanism effectively creates a functional giant component in the routing dynamics, even when the underlying structural connectivity is degraded. The correlation between a rising max hops and a rising improvement of packet loss rate below $P_c$ indicates that the reverse-flow mechanism efficiently leverages the expanding connected component. Beyond $P_c$, the decline of both parameters marks network fragmentation. Ultimately, this newly established relationship bridges the static geometric theory of percolation and the dynamic theory of network forwarding, demonstrating that topological symmetry can induce a functional routing transition in network dynamics that is analogous to, but distinct from, the structural phase transition in percolation. This insight enriches our understanding of resilience in structured networks, positioning symmetry not just as a structural property but as a fundamental enabler of adaptive, fault-tolerant communication.

The challenge of efficient routing using only local information has been a central topic in network theory [35]. In comparison with existing fault-tolerant routing schemes, this work is contextualized against two representative approaches: convergence-free IP routing, exemplified by FCP, and stateless SDN data plane traversal methods. FCP embeds failure information in packet headers to enable distributed path repair, eliminating convergence but incurring per-packet overhead. SDN-based traversals (e.g., random walk or DFS) enable controller-independent forwarding but often introduce path stretch and rule complexity. While both aim to improve fault reachability, FCP requires global network visibility and in-packet state, while SDN traversals rely on flow tables or tags—both implying protocol support or state maintenance.

In contrast, the proposed topology-oblivious forwarding mechanism exploits rotational and reflective symmetries of 2D torus topologies to achieve fault bypass via a localized reverse-flow strategy — without protocol modifications, additional headers, or control-plane dependence. It avoids the in-packet overhead of FCP and the flow-table dependency of SDN, achieving protocol independence and zero control overhead. Unlike protocol-agnostic approaches that rely on flat labeling schemes like ROFL [36], name-based forwarding [37] or in-band failure distribution like FCP, our mechanism requires no modification to packet headers or addressing schemes, leveraging only the inherent topological symmetry. Notably, this work reveals a phase transition in reverse-flow dynamics and establishes a link to percolation theory, reducing packet loss by up to 17.5% in a 16×16 torus while offering new theoretical insights into symmetry-driven self-organization. Thus, it provides both a lightweight fault-tolerance mechanism and an interdisciplinary advance linking network dynamics to statistical physics, with implications for future high-performance networks such as satellite constellations and quantum networks.

## VI. Conclusion

This paper has introduced a lightweight, fault-tolerant forwarding mechanism that exploits the intrinsic symmetries of toroidal networks to enable asynchronous packet delivery without control-plane dependency. By modeling packet flow via a topological potential gradient and strategically harnessing failure-induced reverse flows, our approach guides packets around faults efficiently. The proposed RF-CF and RF-LF strategies ensure packets converge to forward-flow phase transition points, guaranteeing reachability under dynamic link and node failures. Extensive simulations on a 16×16 torus demonstrate a packet loss reduction of up to 17.5%, with the RF-LF strategy responsible for 28% of successful deliveries. Beyond performance gains, this work uncovers a profound connection between the dynamic phase transition of reverse-flow routing and the static structural transition in percolation theory, bridging network dynamics with statistical physics. By transforming topological symmetry into communication reliability, our method provides a protocol-agnostic primitive to enhance the resilience of distributed learning, optimization, and inference tasks in next-generation networks, including satellite constellations and quantum networks..

## Appendix

*A. Pseudocode*

(1) Reverse Flow Generation and Transmission

When packet arrives at node A with destination D:
1. Retrieve primary egress interface E for D from routing table
2. Check link status of E:
   if E is UP then
     Forward packet normally via E
   else
     // Reverse flow triggered
     Let k = degree (node A)
     if k == 2 then
       Let O = opposite interface of incoming link
       if O is UP then
         Forward via O // Creates reverse flow
       else
         Drop packet
       end if
     else if k == 4 then
       if policy == OPPOSITE_FIRST then
         if opposite interface O is UP then
           Forward via O
         else
           for each side interface S in [S1, S2] do
              if S is UP then
                 Forward via S
                 break
              end if
           end for
           if no side interface UP then
              Drop packet
           end if
         end if
       else if policy == SIDE_FIRST then
         for each side interface S in [S1, S2] do
           if S is UP then
              Forward via S
              break
           end if
         end for
         if no side interface UP then
           if opposite interface O is UP then
              Forward via O
           else
              Drop packet
           end if
         end if
       end if
     end if
   end if

(2) Reverse Flow Forwarding and Annihilation

When reverse flow packet arrives at node B via interface $I_{in}$:
1. Retrieve primary egress interface E for destination D
2. Compare $I_{in}$ and E:
   if $I_{in}$ == E then
     // Packet is reverse flow
     Let k = degree (node B)
     if k == 2 then
       Let O = opposite interface of $I_{in}$
       if O is UP then
         Forward via O
       else
         Bounce back via $I_{in}$ // Prevent deadlock
       end if
     else if k == 4 then
       if current_policy == OPPOSITE_FIRST then
         if opposite interface O is UP then
           Forward via O
         else
           for each side interface S in [S1, S2] do
              if S is UP then

```
                Forward via S
                break
            end if
        end for
        if no side interface UP then
            Bounce back via I_in
        end if
    end if
    else if current_policy == SIDE_FIRST then
        for each side interface S in [S1, S2] do
            if S is UP then
                Forward via S
                break
            end if
        end for
        if no side interface UP then
            if opposite interface O is UP then
                Forward via O
            else
                Bounce back via I_in
            end if
        end if
    end if
end if
else
    Forward normally via E // Not reverse flow
end if
```

(3) Oscillation Suppression Strategy

Let T = 2 × geodesic_diameter // Threshold for policy switching
For each reverse flow packet:
1. Let H = number of hops traversed
2. if H > T then
    // Switch policy to avoid oscillation
    if current_policy == OPPOSITE_FIRST then
        current_policy = SIDE_FIRST
    else
        current_policy = OPPOSITE_FIRST
    end if
    Reset hop count H = 0
3. Continue processing with current_policy

*B. Performance on Different Size Topology*

This appendix provides a comprehensive analysis of the performance of our proposed reverse-flow mechanism on torus topologies of different sizes: 8×8, 10×10, 12×12, 14×14, and 16×16. The results are presented in Fig. B-1 for bond percolation and Fig. B-2 for site percolation. Each figure contains panels (a)-(e) showing the packet loss improvement of different methods (LFA, RF-CF, RF-LF) over the Normal Flow (NF) baseline, panels (f)-(h) summarizing the trend of this improvement across topologies, and panels (i)-(j) showing the Reverse-Flow (RF) packet ratio.

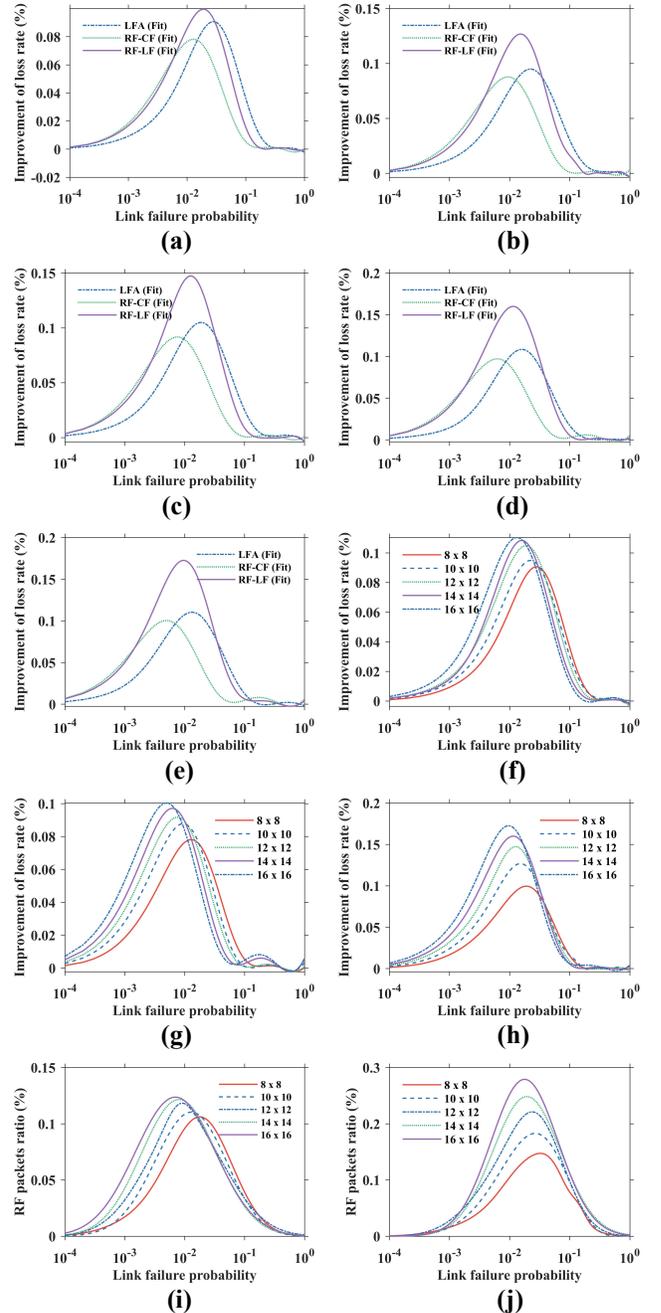

**Fig. B-1.** Packet loss improvement of different methods on (a) 8x8 topology (b) 10×10 topology (c) 12×12 topology (d) 14×14 topology (e) 16×16 topology. Pakcet loss improvement of (f) LFA method (g) RF-CF method (h) RF-LF method on different topology. RF packets ratio of (i) RF-CF method (j) RF-LF method on different topology.

As seen in Figs. B-1(a-e) and B-2(a-e), the qualitative behavior of all methods is consistent across network sizes. The RF-LF strategy consistently outperforms both RF-CF and LFA across all failure probabilities and for all network scales. The absolute packet loss reduction is most significant around the critical failure probability $P_c$ for each topology. Figs. B-1(f-h) and B-2(f-h) consolidate the maximum packet loss improvement achieved by each method against network size. The RF-LF strategy (Figs. B-1h, B-2h) reliably provides the highest improvement. Under bond percolation, its maximum improvement ranges from approximately 22% in the 8×8 torus to 17.5% in the 16×16 torus. The improvement generally decreases as the network grows larger. This is expected because in a larger network, the proportional impact of a single link failure on end-to-end connectivity is slightly lower, and packets have more alternative paths even without the reverse-flow mechanism. However, the fact that significant improvement persists in the 16×16 network demonstrates the method's effectiveness in large-scale deployments.

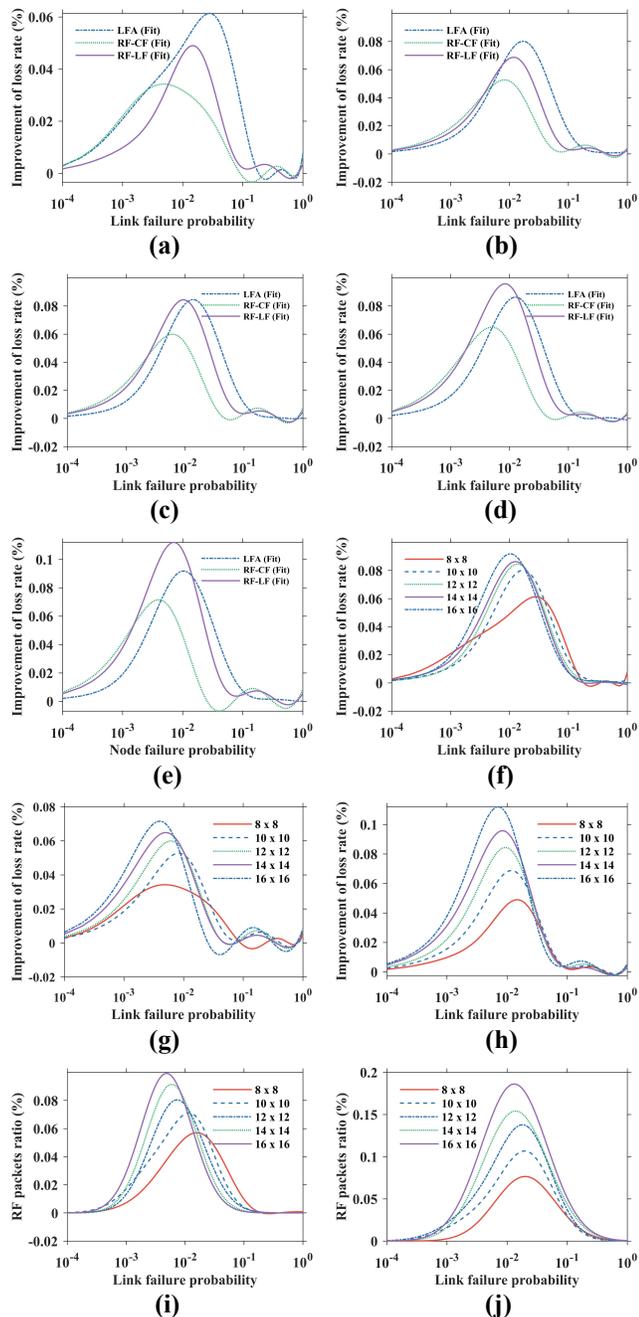

**Fig. B-2.** Packet loss improvement of different methods on (a) 8×8 topology (b) 10×10 topology (c) 12×12 topology (d) 14×14 topology (e) 16×16 topology. Pakcet loss improvement of (f) LFA method (g) RF-CF method (h) RF-LF method on different topology. RF packets ratio of (i) RF-CF method (j) RF-LF method on different topology.

The RF packet ratio (Figs. B-1(i-j), B-2(i-j)) quantifies the proportion of successfully delivered packets that leveraged the reverse-flow mechanism at least once. This metric reveals the direct contribution of our mechanism to reliable delivery. The RF-LF strategy consistently achieves a higher RF packet ratio than RF-CF, explaining its superior performance in loss reduction. Under bond percolation (Fig. B-1i,j), the RF-LF ratio peaks between 25%-30%, while RF-CF peaks between 12%-15%. This indicates that the lateral-facing strategy is more effective at creating useful detours that reconnect to forward paths. The ratio trends with failure probability, peaking near $P_c$ where the network is critically connected and intelligent bypasses are most valuable. Beyond $P_c$, the ratio drops as network fragmentation makes successful delivery via any means increasingly rare.

## References

[1] B. Guo et al., "Resilience of Mega-Satellite Constellations: How Node Failures Impact Inter-Satellite Networking Over Time?," in IEEE Transactions on Communications, doi: 10.1109/TCOMM.2025.3610221..

[2] A. Kubica and M. Vasmer, "Single-shot quantum error correction with the three-dimensional subsystem toric code," Nat. Commun., vol. 13, p. 6272, 2022. doi: 10.1038/s41467-022-33906-5..

[3] A. Nedic and A. Ozdaglar, "Distributed subgradient methods for multi-agent optimization," IEEE Transactions on Automatic Control, vol. 54, no. 1, pp. 48-61, Jan. 2009.

[4] T. Lan, D. Zhou, M. Sheng and J. Li, "Capacity Analysis of LEO Mega-Constellations With Quasi-Torus Topologies," in IEEE Transactions on Communications, vol. 73, no. 9, pp. 8198-8210, Sept. 2025, doi: 10.1109/TCOMM.2025.3547752.

[5] G. Q. Mao, B. D. O. Anderson, and B. Fidan, "Path loss exponent estimation for wireless sensor network localization," Computer Networks, vol. 51, no. 10, pp. 2467-2483, Jul. 2007.

[6] A. Reisizadeh et al., "FedPAQ: A Communication-Efficient Federated Learning Method with Periodic Averaging and Quantization," in Proceedings of the 23rd International Conference on Artificial Intelligence and Statistics, 2020.

[7] A. Tootoonchian et al., "On Controller Performance in Software-Defined Networks," in Proceedings of the 2nd USENIX Conference on Hot Topics in Management of Internet, Cloud, and Enterprise Networks and Services, 2012.

[8] C. Xian, Y. Zhao, G. Wen, and G. Chen, "Robust Event-Triggered Distributed Optimal Coordination of Heterogeneous Systems Over Directed Networks," IEEE Transactions on Automatic Control, vol. 69, no. 3, pp. 1234-1249, Mar. 2024.

[9] A. Markopoulou, G. Iannaccone, and S. Bhattacharyya, "Characterization of failures in an operational IP backbone network," IEEE/ACM Transactions on Networking, vol. 16, no. 4, pp. 749-762, Aug. 2008.

[10] M. Caesar et al., "Dynamic route recomputation considered harmful," ACM SIGCOMM Computer Communication Review, vol. 40, no. 2, pp. 66-71, Apr. 2010.

[11] K. Lakshminarayanan, M. Caesar, and I. Stoica, "Achieving convergence-free routing using failure-carrying packets," in Proceedings of the 2007 conference on Applications, technologies, architectures, and protocols for computer communications, 2007, pp. 241-252.

[12] N. Katta, H. Zhang, and J. Rexford, "Infinite CacheFlow in Software-Defined Networks," in Proceedings of the third workshop on Hot topics in software defined networking, 2014, pp. 175-180.

[13] A. Gopalan and S. Ramasubramanian, "IP Fast Rerouting for Multi-Link Failures," IEEE/ACM Transactions on Networking, vol. 24, no. 5, pp. 3014-3025, Oct. 2016.

[14] M. Borokhovich, L. Schiff, and S. Schmid, "The show must go on: Fundamental data plane connectivity services for dependable SDNs," Computer Communications, vol. 116, pp. 172-183, Jan. 2018.
[15] W. J. Dally and B. Towles, "Route packets, not wires: On-chip interconnection networks," in Proceedings of the 38th Design Automation Conference, 2001, pp. 684-689.
[16] J. Duato, S. Yalamanchili, and L. Ni, Interconnection Networks: An Engineering Approach. San Francisco, CA, USA: Morgan Kaufmann, 2003.
[17] Q. Chen, T. Wang, and Y. Yang, "3-ISL Topology: Routing Properties and Performance in LEO Megaconstellation Networks," IEEE Transactions on Aerospace and Electronic Systems, vol. 61, no. 6, pp. 4961-4972, Dec. 2025.
[18] Q. Chen, T. Wang, and Y. Yang, "Shortest Path in LEO Satellite Constellation Networks: An Explicit Analytic Approach," IEEE Journal on Selected Areas in Communications, vol. 42, no. 4, pp. 1175-1187, Apr. 2024.
[19] Y. Yang, Y. Liu, and W. M. Dai, "Recursive diagonal torus: An interconnection network for massively parallel computers," IEEE Transactions on Parallel and Distributed Systems, vol. 12, no. 7, pp. 701-715, Jul. 2001.
[20] M. Besta and T. Hoefler, "Slim Fly: A cost effective low-diameter network topology," in Proceedings of the International Conference for High Performance Computing, Networking, Storage and Analysis, 2014, pp. 348-359.
[21] S. Wehner, D. Elkouss, and R. Hanson, "Quantum internet: A vision for the road ahead," Science, vol. 362, no. 6412, p. eaam9288, Oct. 2018.
[22] A. K. Ekert, "Quantum cryptography based on Bell's theorem," Physical Review Letters, vol. 67, no. 6, pp. 661-663, Aug. 1991.
[23] Z. Zhang, L. Ma, K. K. Leung, and F. Le, "More is not always better: An analytical study of controller synchronizations in distributed SDN," IEEE/ACM Trans. Netw., vol. 29, no. 4, pp. 1580–1590, 2021.
[24] H. Li, G. Shou, Y. Hu, and Y. Liu, "SDN/NFV enhanced time synchronization in packet networks," IEEE Syst. J., vol. 15, no. 4, pp. 5634–5645, 2021.
[25] B. Yang, J. Liu, S. Shenker, J. Li, and K. Zheng, "Keep forwarding: Towards k-link failure resilient routing," in Proc. IEEE INFOCOM, 2014, pp. 1617–1625.
[26] M. Chiesa et al., "On the resiliency of randomized routing against multiple edge failures," in Proc. Int. Colloq. Autom., Lang., Program., 2016, pp. 1–15.
[27] Y. Yang, M. Xu, and Q. Li, "Fast rerouting against multi-link failures without topology constraint," IEEE/ACM Trans. Netw., vol. 26, no. 1, pp. 384–397, 2018.
[28] G. Barrenetxea, B. Berefull-Lozano, and M. Vetterli, "Lattice networks: Capacity limits, optimal routing, and queueing behavior," IEEE/ACM Trans. Netw., vol. 14, no. 3, pp. 492–505, 2006.
[29] Q.-P. Gu and S. Peng, "Fault tolerant routing in toroidal networks," IEICE Trans. Inf. Syst., vol. E79-D, no. 8, pp. 1153–1159, 1996.
[30] A. Bossard and K. Kaneko, "Cluster-fault tolerant routing in a torus," Sensors, vol. 20, no. 11, p. 3286, 2020.
[31] V. Puente et al., "Adaptive bubble router: A design to improve performance in torus networks," in Proc. Int. Conf. Parallel Process., 1999, pp. 58–67.
[32] K. Menger, "Zur allgemeinen Kurventheorie," Fundam. Math., vol. 10, pp. 96–115, 1927.
[33] X. Lv and L. Chen, "A potential-based clustering routing protocol for wireless sensor network," in Proc. Int. Conf. Comput. Problem-Solving, 2012, pp. 151–154.
[34] S. R. Broadbent and J. M. Hammersley, "Percolation processes: I. Crystals and mazes," Proc. Camb. Phil. Soc., vol. 53, pp. 629–641, 1957.
[35] J. M. Kleinberg, "Navigation in a small world," Nature, vol. 406, p. 845, 2000.
[36] A. Duch et al., "ROFL: Routing on flat labels," ACM SIGCOMM Comput. Commun. Rev., vol. 36, no. 4, pp. 363–374, 2006.
[37] V. Jacobson et al., "Networking named content," in Proc. ACM CoNEXT, 2009, pp. 1–12.

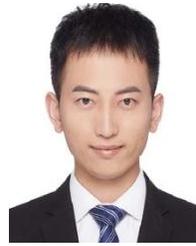

**Shenshen Luan** (Member, IEEE) received the B.S. degree in 2016 and the M.S. degree in 2019 in Electronics Science and Technology from Electronics and Information Engineering School, Beihang University, Beijing, China. He is now pursuing his Ph.D. degree in Beihang University. His main research focuses on satellite network, electromagnetic imaging and wireless communication. His interests include artificial intelligence, microwave sensing and optimization algorithms.

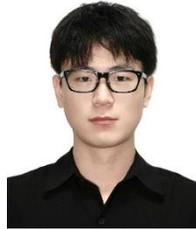

**Yumo Tian** received the B.S. and Ph.D. degrees in Electronics Science and Technology from Electronics and Information Engineering School, Beihang University, Beijing, China, in 2019 and 2025. His main research focuses on microwave photonics, electro-optical sensing, and photonic integration. His interests include electro-optical sensors and thin-film lithium niobate photonic integration.

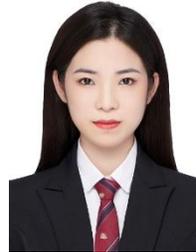

**Xinyu Zhang** received the M.S. degree in Electronics Science and Technology from the University of Chinese Academy of Sciences, Beijing, China, in 2023, and she is currently pursuing the Ph.D. degree in Electronics Science and Technology from Electronics and Information Engineering School, Beihang University, Beijing, China. Her main research interests include electric field measurement techniques and single-channel blind source separation.

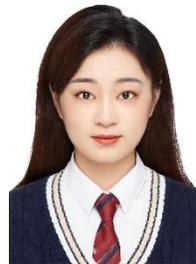

**Qingwen Zhang** received the B.S. degree in Communication Engineering from Beijing Jiaotong University, Beijing, China, in 2023, and she is currently pursuing the M.S. degree in Electronics Science and Technology from the Electronics and Information Engineering School, Beihang University, Beijing, China. Her main research interests include electric field measurement techniques and drone recognition.

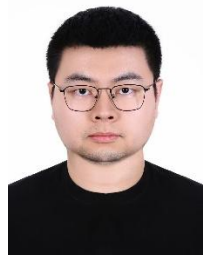

**Tianheng Wang** received the Ph.D. degree in Electronic Information Engineering from Beihang University, Beijing, China, in 2024. He is currently an engineer in Hefei Innovation Research Institute of Beihang University. His research interests include microwave photonics sensing, optical signal processing, and broadband microwave imaging.

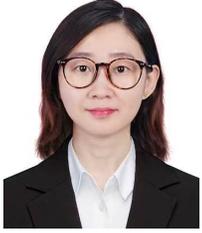

**Yan Yang** received the Ph.D. degree in Circuits and Systems from Beihang University, Beijing, China, in January 2022. Since 2022, she has been a postdoctoral researcher with Information and Communication Engineering Postdoctoral Research Station, Beihang University. Her research focuses on ultra-wideband electromagnetic signal sensing, high-precision measurement techniques, and real-time signal processing.

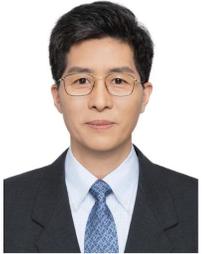

**Shuguo Xi**e (Member, IEEE) received the M.S. and Ph.D. degrees in radio physics from Wuhan University, Wuhan, China, in 1992 and 2001, respectively. He started his work at Wuhan University, in 1992. In 1996, he was a Visiting Researcher with the Laboratoire de Physique et Chimie de l'Environnement, CNRS, Toulouse, France. He joined Beihang University, Beijing, China, in 2006, where he is currently a Professor with the School of Electronic Information Engineering. His research interests include radio wave propagation, antenna design, electromagnetic compatibility and its measurement, and macro modeling of electronic equipment.